# A Distributed Shared Memory Model and C++ Templated Meta-Programming Interface for the Epiphany RISC Array Processor

David Richie[1], James Ross[2], and Jamie Infantolino[2]

[1]*Brown Deer Technology, MD, USA*
drichie@browndeertechnology.com
[2]*U.S. Army Research Laboratory, MD, USA*
{james.a.ross176.civ, jamie.k.infantolino.civ}@mail.mil

**Abstract**
The Adapteva Epiphany many-core architecture comprises a scalable 2D mesh Network-on-Chip (NoC) of low-power RISC cores with minimal uncore functionality. Whereas such a processor offers high computational energy efficiency and parallel scalability, developing effective programming models that address the unique architecture features has presented many challenges. We present here a distributed shared memory (DSM) model supported in software transparently using C++ templated meta-programming techniques. The approach offers an extremely simple parallel programming model well suited for the architecture. Initial results are presented that demonstrate the approach and provide insight into the efficiency of the programming model and also the ability of the NoC to support a DSM without explicit control over data movement and localization.

*Keywords:* 2D RISC array, Adapteva Epiphany, Parallella, distributed shared memory, templated meta-programming, parallel programming model

## 1 Introduction

The emergence of new and rapidly changing parallel processor architectures brings challenges for software development specifically directed toward portability and performance. Whereas software developers may have a mix of well-established, or newly developed, portable programming APIs available for a given architecture, they are rarely sufficient to provide optimum performance without significant effort, refactoring and tuning with source-level architecture-specific code. The result is a fragmentation of the source-code base supporting a given application, or the development of architecture-specific implementations of entire applications. Although portability is easy to establish in an API specification, enabling performance-portable code remains a significant challenge that dominates efforts to utilize the latest parallel processor architectures.

The development of solutions for performance-portable code remains an open challenge of great interest in computer science as it is applied to high-performance computing. At issue is not the ability to achieve the maximum theoretical performance for every algorithm comprising a given software package, since this will always require heroic efforts and some degree of architecture-specific customization of software. At present, it is proving difficult to achieve even relatively good performance measured against the capabilities of a given parallel architecture. In some cases, non-portable code is required regardless of performance objectives. The Epiphany processor architecture has provided an example of the challenges faced in parallel programmability that must be addressed to support performance-portable code.

The Adapteva Epiphany RISC array architecture [1] is a scalable 2D array of low-power RISC cores with minimal un-core functionality supported by an on-chip 2D mesh network for fast inter-core communication. The Epiphany-III architecture is scalable to 4,096 cores and represents an example of an architecture designed for power-efficiency at extreme on-chip core counts. Processors based on this architecture exhibit good performance/power metrics [2] and scalability via 2D mesh network [3][4], but require a suitable programming model to fully exploit the architecture. A 16-core Epiphany-III processor [5] has been integrated into the Parallella mini-computer platform [6] where the RISC array is supported by a dual-core ARM CPU and asymmetric shared-memory access to off-chip global memory. We have recently published results for threaded MPI [7], an OpenSHMEM programming model for Epiphany [8][9], a hybrid programming model [10], and other advances in runtime performance and interoperability [11].

RISC array processors, such as those based on the Epiphany architecture, may offer significant computational power efficiency in the near future with requirements in increased core counts, including long-term plans for exascale platforms. The power efficiency of the Epiphany architecture has been specifically identified as both a guide and prospective architecture for such platforms [12]. The Epiphany-IV processor has a performance efficiency of 50 GFLOPS/W [2] making it one of the most efficient parallel processors based on general-purpose cores and satisfying the threshold for exascale computing with a power budget of 20 megawatts [13]. This architecture has characteristics consistent with future processor predictions of hundreds [14] and thousands [15][17] of cores on a chip. The 1024-core, 64-bit Epiphany-V was recently taped out and is anticipated to have much higher performance and energy efficiency [17].

We present an investigation into the use of C++ templated meta-programming (TMP) techniques for data layout and parallel loop order abstraction as a parallel programming API targeting the Epiphany architecture. Using this API we explore a transparent distributed shared memory (DSM) model for Epiphany that eliminates the need to manage local data movement between cores. The performance of the model using various configurations provides interesting benchmark data for the analysis of the performance characteristics of the Epiphany 2D mesh network. The parallel programing model presented here is portable and was initially developed for other processor architectures. The successful application of this parallel programming model to the Epiphany architecture, especially given the significant architectural constraints and differences from a conventional multi-core CPU, provide an interesting validation of the approach. There is no special treatment of the Epiphany architecture in application code. Specifically, there are no co-processor offload semantics or host/co-processor co-design used in this programming model, which distinguishes it from many of the early attempts at programming the Epiphany processor. All architecture-specific issues are handled within a TMP backend just as they are addressed for other supported platforms.

Our main contributions are as follows: we present a parallel programming API based on C++ templated meta-programming techniques for data layout and loop order abstraction, we apply these techniques with a backend for the Epiphany architecture implemented to support a DSM model, and we use the various configurations available within the API to explore the performance of the Epiphany 2D mesh network for supporting this memory model. Section 2 describes the relevant features of the Epiphany RISC array architecture and the motivation for the proposed memory model and API. Section

3 describes the portable C++ TMP techniques and their application to the Epiphany architecture. Section 4 describes the transparent support for a DSM model implemented within the TMP backend. Section 5 describes the application to a simple n-body benchmark and a more complex finite-difference time-domain (FDTD) solver. Section 7 provides concluding observations.

## 2   Background

The Adapteva Epiphany MIMD architecture is a scalable 2D array of RISC cores with minimal uncore functionality connected with a fast 2D mesh Network-on-Chip (NoC). Figure 1 shows the high-level architectural features of the processor. Each of the 16 Epiphany-III mesh nodes contains 32 KB of shared local memory (used for both program instructions and data), a mesh network interface, a dual-channel DMA engine, and a RISC CPU core. Each RISC CPU core contains a 64-word register file, sequencer, interrupt handler, arithmetic logic unit, and a floating point unit. Each processor tile is very small at 0.5 mm$^2$ on the 65 nm process and 0.128 mm$^2$ on the 28 nm process. Peak single-precision performance for the Epiphany-III is 19.2 GFLOPS with a 600 MHz clock. Fabricated on the 65 nm process, the Epiphany-III consumes 594 mW for an energy efficiency of 32.3 GFLOPS per watt. The 64-core Epiphany-IV, fabricated on the 28 nm process, has demonstrated energy efficiency exceeding 50 GFLOPS per watt [2], and 1024-core 64-bit Epiphany-V recently taped out on 16 nm is anticipated to be much higher [16].

The Epiphany architecture is based on a 2D array of low-power 32-bit RISC cores, each with 32 KB of fast local memory and a robust mesh network for fast inter-core communication. The fully memory-mapped architecture allows shared memory access to global off-chip memory and shared non-uniform memory access to the local memory of each core. A block diagram of the Epiphany architecture is shown in Figure 1.

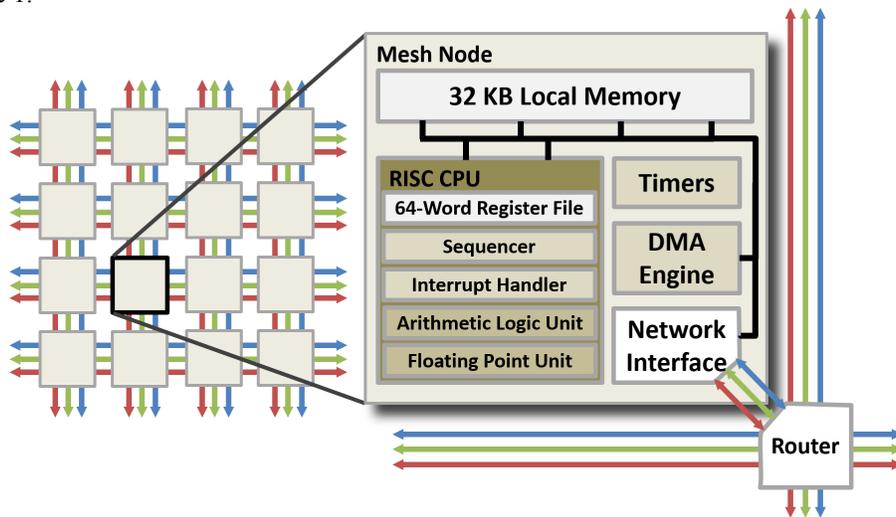

**Figure 1:** The Epiphany RISC array architecture. RISC cores are connected through a point-to-point network for signaling and data transfer. Communication latency between cores is low, but the amount of addressable data contained on a mesh node is low (32 KB). The three networks shown above handle local read transactions, local write transactions, and off-chip memory transactions.

Each core in the RISC array has only 32 KB that must be used for both instructions and local data storage. Although each core has shared memory access to off-chip global memory, this access is significantly slower than local memory. The architectural feature that must be exploited in any implementation is the extremely low latency access to local memory between cores supported by the 2D mesh network.

The present work is motivated by prior investigations into developing a parallel programming model for the Epiphany architecture, including OpenCL [10], threaded MPI [7], and OpenSHMEM [8][9] support for the Epiphany architecture. In all cases the parallel programming model involved explicit data movement between the local memory of each core in the RISC array, or to/from the off-chip global DRAM. The absence of a hardware cache necessitated this movement to be controlled in software explicitly. Also relevant to the present work, progress was made in the development of a more transparent compilation and run-time environment whereby program binaries could be compiled and executed directly on the Epiphany co-processor of the Parallella platform without the use of an explicit host offload co-design model [11]. This latter development led directly to the exploration of compiling more complex code using C++ rather than the more limited use of C. Given the significant constraints in local memory size, and the practical limit of approximately 16KB for program instructions (leaving 16KB for data storage), the ability to compile C++ code for Epiphany remained an open question in itself, since it was unclear whether the overhead would be too great.

The need for compiling highly efficient and optimized code is critical in high-performance computing applications, and in the context of C++, requires more sophisticated techniques than standard object-oriented programming which incurs too high of a performance penalty for the convenience of abstractions. The use of templated meta-programming (TMP) was introduced to resolve this issue, and has been utilized in many C++ packages designed for object oriented programming. These techniques provide abstraction and at the same time are able to generate highly efficient code through partial template specialization at compile time. The use of such techniques for developing performance-portable code is an active area of computer science research. In this work we apply a TMP package, CLETE-2, developed for data layout and parallel loop order abstraction, to enable a simple parallel programing model with support for a transparent DSM model for the Epiphany architecture. In this model, each parallel thread is able to transparently access shared arrays physically distributed over the local memory of all cores within the RISC array, using a conventional C++ array accessor. The physical data layout as well as parallel loop order is determined by a compile-time type-selected configuration (nothing more than a C++ typedef) abstracted from the actual application code.

## 3   C++ Templated Meta-Programming: CLETE-2

Templated meta-programming was originally applied to expression templates as a solution to the abstraction penalty of object-oriented programming. At issue was the use of overloaded operators for constructing expressions with containers, e.g., arrays, which in a conventional OO implementation incur a significant abstraction penalty. The solution of encoding container expressions as a compile-time object of complex type (expression template) revealed a powerful technique for performing *compile-time code transformations*. Solution was clever, and applied C++ template partial specialization in ways never envisioned by the original standard development. This technique is now generally applied to produce high-performance C++ code in various software packages and applications. In this work we utilize TMP techniques to separate an abstract representation of data arrays and parallel loops from any specific implementation of physical data layout and parallel work decomposition. The approach includes support for expression templates as well as multi-expression lambda functions required for encapsulation of more complex computational kernels, not expressible as a single equation.

An important aspect of the approach described below is the very deliberate, rigorous adherence to the C++ standard, requiring no language extensions, additional tools, or backend libraries, beyond those required to support ordinary C++ programming of the target architecture. Although it is tempting to introduce non-standard language extensions, or augment an ordinary compiler with special tools, in order to solve the latest programming challenge introduced by a new parallel architecture, this undermines long-term portability and stability of developed applications. Support for new architectures should be implemented within prevalent language standards and compilers that are more than sufficient for expression and compilation of parallel code.

TMP in some sense builds a layer between the application code and the compiler, and any low level optimizations the latter is capable of performing, so that much higher level code transformations can be performed to better enable the compiler optimizations and perform optimizations beyond the capabilities of the compiler. We use a package called CLETE (Compute Layer Expression Template Engine) version 2, which is under active development as a portable abstraction of data layout and parallel loop ordering. The package is comprised entirely of header files and has no additional backend tools or libraries. CLETE was originally developed to support accelerator offload as an application of expression templates, and is based on PETE.

Data layout is abstracted by using containers for multi-dimensional multi-component arrays in which both the data type and data layout may be specified as templated type parameters. C++ variadic templates are used to construct a layout generator to provide the programmer with a free-form specification that is easier to understand and use, and does not require specific ordering of template parameters. As a simple example in which an abstract container is created for a two-dimensional array of points (x,y, and z) of scalar type float,

```
typedef ArrayType<float,2> array2d_t;
LayoutSpec<
      MultiArray< array2d_t, 3 >, [optional parameters]
>::layout_t points_t;
points_t points(size1, size2);
layout_compnent<1, points_t>::type_t x(points);
layout_compnent<1, points_t>::type_t y(points);
layout_compnent<1, points_t>::type_t z(points);
```

The result is the allocation of storage for 3 logical arrays, x, y, z, such that the actual layout is determined by the type-selected options. In all usage, the elements of the three arrays are accessed using a standard accessor, e.g., x(i,j). More complex data layouts for arrays are enabled by a range of optional template parameters. Examples of some of optional parameters are shown in Table 1, and which may be specified in any order making the syntax for data layout convenient for the programmer.

**Table 1:** Optional parameters controlling data layout

| Optional Template Parameter | Effect |
|---|---|
| `Group<>` | Control the grouping of quantities |
| `IndexOrder<>` | Control the index ordering of the accessor |
| `Tile<>` | Control the tiling of multi-dimensional arrays, may be nested |
| `Part<>` | Controls the transparent partitioning of arrays, may be nested |

The arrays defined above fully support expression template encapsulation and the use of interval objects similar to the style used in POOMA. For example, the following code fragment would perform the trivially parallel outer product on x and y, where actual loops are abstracted and the code instantiated at compile-time can be parallelized or vectorized transparently,

```
Interval I(size1);
Interval J(size2);
z(I, J) = x(I, J) * y(I, J)
```

More complicated computational kernels, not expressible as a single expression, can be treated using a `parallel_for<>` construct using C++ lambda functions, with expression templates nested within. As an example, the following code fragment will update x and y based on values of z within the same abstracted loop,

```
Interval I1(1, size1);
Interval J1(1, size2);
parallel_for( I1, J1, [&] ( auto& I, auto& J ) {
      x(I, J) = z(I, J) + z(I, J-1);
      y(I, J) = z(I, J) + Z(I-1, J);
});
```

Notice the introduction of shifts on the array indexing which may be used, e.g., in stencil operations where values are updated using current values of neighboring points. Using syntax similar to that used for specifying data layout, the specific code transformations to be applied to loop constructs may be specified with a plan generator. In the example code fragment below, `plan_t` specifies that an OpenMP parallelization should be used for the parallel loop.

```
typedef PlanSpec< array2d_t, Model<OpenMP> >::Plan_t plan_t;
Interval I1(1, size1);
Interval J1(1, size2);
parallel_for<plan_t>( I1, J1, [&] ( auto& I, auto& J ) {
      x(I, J) = z(I, J) + z(I, J-1);
      y(I, J) = z(I, J) + Z(I-1, J);
});
```

The application of the CLETE-2 package requires a compiler that correctly implements the C++17 standard specification and also correctly optimizes C++ template partial specializations to produce efficient code. In this work we utilize the GCC 5.4 complier for targeting the Epiphany processor. We additionally rely on the COPRTHR-2 SDK which provides run-time support for the Epiphany processor including support for fast SPMD direct co-processor execution, without requiring offload semantics or co-design with the ARM CPU on the Parallella platform. As a result, the compilation and run-time environment used in this work resembles that of an ordinary Linux platform with a multi-core processor.

## 4 Distributed Shared Memory Model

The key architectural feature exploited with the software DSM design is the fast on-chip mesh network that enables low-latency access to the local memory of any core. Each core is fully memory-mapped in the global address space, allowing load and store operations by dereferencing global pointers. A global pointer to any remote core may be trivially calculated based on the core identification within the RISC array. Notably, remote writes are significantly faster than remote reads. Prior work emphasized one-sided remote writes to achieve higher performance.

Each core has a flat symmetric local memory map and also a global address map which is translated by the network router. A global address may be directly dereferenced within software and the operation completes transparently to user code. Local allocations in memory may be physically contiguous, however, a distributed symmetric allocation on all cores does not have the corresponding contiguous global addressing. There is no hardware functional unit for mapping globally addressable memory segments to a contiguous virtual memory range. Although translating local address to global address is trivially handled by the local-to-global address translation subroutine, the additional software development overhead for explicitly tracking pointers and indices complicates the use of an array allocated with storage distributed over the local memory of multiple cores. With the use of C++ TMP techniques described above, the necessary pointer calculations can be abstracted using a common array accessor, and of equal important, made efficient, enabling implicit indexing to distributed shared memory.

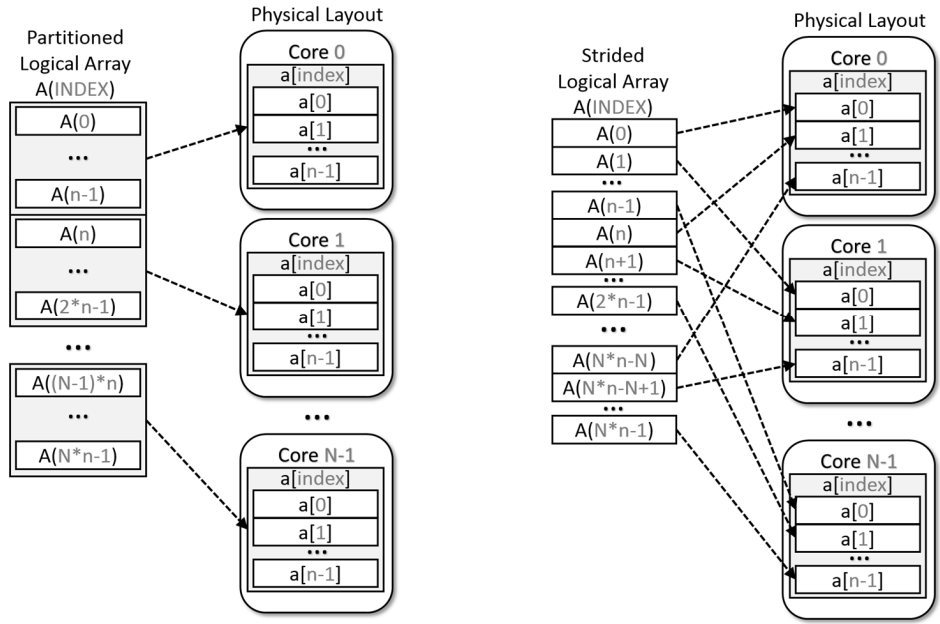

**Figure 2:** Logical arrays may be physically distributed in a number of ways. On the left, the logical array is partitioned with a contiguous physical layout distributed amongst multiple cores. On the right, a strided pattern is used for the physical layout. Each layout has different performance optimization tradeoffs.

A logical array within C++ can have a number of physical layouts, but we've chosen to look at two examples: partitioned and strided arrays. A partitioned array distributes a contiguous portion of the logical array to each contiguous portion of symmetric memory with the cores. The strided array distributes the memory in a round-robin manner so that each consecutive logical index is stored on a different core.

## 5 Applications

We evaluated our DSM and C++ TMP approaches with a simple n-body benchmark and a more complex finite-difference time-domain (FDTD) solver. In the case of the n-body benchmark a simple code is used with a straightforward implementation, and without optimizations. An equivalent reference C code is used to compare the efficiency of the code generate using the C++ TMP techniques. For the case of the FDTD application, an existing application is used that was developed in prior work to utilize the C++ TMP techniques.

For the n-body application, the serial core performance is compared between the C++ TMP implementation and the reference C code. For 512 particles the performance of the C++ TMP benchmark is 367 MFLOPS, which is slightly better than the equivalent C reference code, indicating that the compiler does a good job optimizing the partial template specializations to produce efficient instructions. The size of the executable binary is critical with the Epiphany architecture since the core-local memory must be shared between instructions and data. The amount of memory available with the

C++ code is only slightly less than that of the C code, being 70% and 74% of the core-local memory, respectively. Enabling DSM and configuring a parallel execution plan, the performance of the C++ code achieves slightly less than 1 GFLOP for 6,144 particles. Note that the DSM is necessary to allow for the larger number of particles because only a single copy is necessary compared to highly optimized code which passes around copies of data. It should be emphasized that the extension to the DSM and the parallel execution are nearly transparent to the programmer, requiring only a change in two typedefs, not requiring a change from the serial application. This simplicity is the benefit for the programmer of the C++ TMP techniques.

The DSM and C++ TMP techniques were then tested using a more complicated application. Specifically, the 2D FDTD application that was previously implemented using the CLETE-2 TMP interface was compiled for the Epiphany architecture using the DSM. The only changes made to the FDTD code were the removal of C++ iostream operations since they are not presently supported for Epiphany. However, the computational kernel was left unchanged from that of previous work targeting more conventional architectures. No difficulties in compilation of the application were encountered. The successful execution of this application on the Epiphany architecture suggests that the overall approach presented here has significant merit and offers a methodology for implementing computational kernels in such a way that data layout and parallel loop order can be extracted and then optimized transparently for very different parallel processor architectures.

# 6 Related Work

The availability of the inexpensive Adapteva Parallella platform with the 16-core Epiphany-III processor has led to several investigations of the architecture and various programming models. We have recently published results for threaded MPI [7], an OpenSHMEM programming model for Epiphany [8][9] and other advances in runtime performance and interoperability [11]. To date, no prior publications have reported on the evaluation of the Epiphany architecture using an object-based software DSM model or high-level C++ TMP techniques for parallel programming. The availability of GCC 5.4 with C++17 support only recently enabled some of the C++ TMP capability.

TMP techniques have been explored using other architectures like CPUs and GPUs [18][19] which lead to the foundation to this work. However, the techniques presented here are uniquely different due to the differences in architectures. Other work has explored templated metaprogramming in different languages and programming environments [20]. We were able to examine this work to improve the implementation for the Epiphany. The work presented here is applicable to the Epiphany chip and other RISC array architectures. As aforementioned, to date there is no previous work that explores these techniques on an architecture like the Epiphany.

# 7 Conclusion

We have demonstrated preliminary results for the core functionality of a software DSM model and C++ TMP techniques for parallel programming are possible within the hardware constraints of the Epiphany architecture. The techniques described in this paper should enable many more software developers to write portable parallel code for the platform. The results allowed us to identify points needed for optimization. The backend must be improved for remote memory access patterns and load balancing. Additionally, a software-defined data cache, configurable by array accessor paramters, may should be beneficial for codes that may reuse remotely addressable memory. We believe these areas may indeed be optimized to enable better performance with the DSM model and C++ TMP techniques on Epiphany-like architectures.

# 8  Acknowledgements

The authors wish to acknowledge the U.S. Army Research Laboratory-hosted Department of Defense Supercomputing Resource Center for its support of this work. This work was supported in part by the Department of Defense (DoD) High Performance Computing Modernization Program (HPCMP) under contract GS04T09DBC0017.

# References

[1]   "Adapteva introduction." [Online]. Available: http://www.adapteva.com/introduction/. [Accessed: 8-Jan-2015].

[2]   A. Olofsson, T. Nordström, and Z. Ul-Abdin, "Kickstarting high-performance energy-efficient manycore architectures with Epiphany," *ArXiv Prepr. ArXiv14125538*, 2014.

[3]   D. Wentzlaff, P. Griffin, H. Hoffmann, L. Bao, B. Edwards, C. Ramey, M. Mattina, C.-C. Miao, J. F. Brown III, and A. Agarwal, "On-chip interconnection architecture of the tile processor," *IEEE Micro*, vol. 27, no. 5, pp. 15–31, Sep. 2007.

[4]   M. B. Taylor, J. Kim, J. Miller, D. Wentzlaff, F. Ghodrat, B. Greenwald, H. Hoffman, P. Johnson, W. Lee, A. Saraf, N. Shnidman, V. Strumpen, S. Amarasinghe, and A. Agarwal, "A 16-issue multiple-program-counter microprocessor with point-to-point scalar operand network," in *2003 IEEE International Solid-State Circuits Conference (ISSCC)*, 2003, pp. 170–171.

[5]   "E16G301 Epiphany 16-core microprocessor," Adapteva Inc., Lexington, MA, Datasheet Rev. 14.03.11.

[6]   "Parallella-1.x reference manual," Adapteva, Boston Design Solutions, Ant Micro, Rev. 14.09.09.

[7]   D. Richie, J. Ross, S. Park, and D. Shires, "Threaded MPI Programming Model for the Epiphany RISC Array Processor," Journal of Computational Science, Volume 9, July 2015, pp. 94–100.

[8]   J. Ross and D. Richie, "Implementing OpenSHMEM for the Adapteva Epiphany RISC array processor," International Conference on Computational Science, ICCS 2016, San Diego, California, USA, 6-8 June 2016

[9]   J. Ross and D. Richie, "An OpenSHMEM Implementation for the Adapteva Epiphany Coprocessor," OpenSHMEM and Related Technologies. Enhancing OpenSHMEM for Hybrid Environments, vol. 10007, pp. 146-159, Dec. 2016, doi:10.1007/978-3-319-50995-2_10

[10]  D. Richie and J. Ross, "OpenCL + OpenSHMEM Hybrid Programming Model for the Adapteva Epiphany Architecture," OpenSHMEM and Related Technologies. Enhancing OpenSHMEM for Hybrid Environments, vol. 10007, pp. 181-192, Dec. 2016, doi: 10.1007/978-3-319-50995-2_12

[11]  D. Richie and J. Ross, "Advances in Run-Time Performance and Interoperability for the Adapteva Epiphany Coprocessor," Procedia Computer Science, vol. 80, Apr. 2016, doi:10.1016/j.procs.2016.05.47

[12]  A. Varghese, B. Edwards, G. Mitra, and A. P. Rendell, "Programming the Adapteva Epiphany 64-Core network-on-chip coprocessor," in *2014 IEEE International Parallel & Distributed Processing Symposium Workshops (IPDPSW '14)*, 2014, pp. 984–992.

[13]  K. Bergman, S. Borkar, D. Campbell, W. Carlson, W. Dally, M. Denneau, P. Franzon, W. Harrod, K. Hill, J. Hiller, and others, "Exascale computing study: Technology challenges in achieving exascale systems," *Def. Adv. Res. Proj. Agency Inf. Process. Tech. Off. DARPA IPTO Tech Rep*, vol. 15, 2008.

[14]  J. Held, J. Bautista, and S. Koehl, "From a few cores to many: A Tera-scale computing research overview," Intel Corporation, White Paper, 2006.

[15]  K. Asanovic, R. Bodik, B. C. Catanzaro, J. J. Gebis, P. Husbands, K. Keutzer, D. A. Patterson, W. L. Plishker, J. Shalf, S. W. Williams, and others, "The landscape of parallel computing research: A view from Berkeley," University of California, Berkeley, Technical Report UCB/EECS-2006-183, 2006.

[16]  S. Borkar, "Thousand core chips: a technology perspective," in *Proceedings of the 44th annual Design Automation Conference (DAC '07)*, 2007, pp. 746–749.

[17]  "Epiphany-V: A 1024-core processor 64-bit System-On-Chip" [Online]. Available: http://www.parallella.org/docs/e5_1024core_soc.pdf. [Accessed: 10-Feb-2017]

[18]  J. Infantolino and D. Richie, "Compile-Time Type Selection of Optimized Data Layout and Memory Access Patterns for FDTD," ACES 2016, Honolulu, HI, USA, 13-17 March 2016, doi: 10.1109/ROPACES.2016.7465383

[19] J. Infantolino, J. Ross and D. Richie, "Portable High-Performance Software Design Using Templated Meta-Programming for EM Calculations," ACES 2017, Florence, Italy, 26-30 March 2017.

[20] Czarnecki, K.; O'Donnell, J.; Striegnitz, J.; Taha, Walid Mohamed (2004). "DSL implementation in metaocaml, template haskell, and C++" (PDF). University of Waterloo, University of Glasgow, Research Centre Julich, Rice University.